# How Graphene Is Cut upon Oxidation?

Zhenyu Li,[1,2,*] Wenhua Zhang,[1,2] Yi Luo,[1,2] Jinlong Yang,[1,*] and Jian Guo Hou[1]

[1]Hefei National Laboratory for Physical Sciences at Microscale, University of Science and Technology of China, Hefei, Anhui 230026, China

[2]Department of Theoretical Chemistry, School of Biotechnology, Royal Institute of Technology, S-10691 Stockholm, Sweden

E-mail: zyli@ustc.edu.cn, jlyang@ustc.edu.cn

Our first principles calculations reveal that oxidative cut of graphene is realized by forming epoxy and then carbonyl pairs. Direct forming carbonyl pair to tear graphene up from an edge position is not favorable in energy. This atomic picture is valuable for developing effective graphene manipulation means. The proposed epoxy pairs may be related to some long puzzling experimental observations on graphene oxide.

Due to its novel physical properties and great potential in various applications, graphene has attracted an intense research interest recently.[i] A big challenge in graphene research is the massive production of high quality samples. The existing physical approaches[ii] prohibit producing and processing graphene on large scales. In this context, the versatile chemistry of carbon may offer a promising alternative for cost-effective mass production of graphene, as demonstrated by its graphene-oxide (GO) synthesis route. Upon oxidation, graphite readily exfoliates as single sheets in water, forming GO. The $\pi$ conjugation in graphene can then be largely restored by reducing GO.[iii,iv] Oxidation now becomes an important chemical means to manipulate graphitic materials.

It has been observed that, during oxidation process, graphitic structures automatically break down into smaller parts.[iv,v] For electronics applications, it is very desirable to cut graphene with designed shape and size. Therefore, an atomistic understanding of the mechanism for such oxidative breakup of graphene sheets is especially valuable. Based on first-principles calculations, an unzipping mechanism has been proposed,[vi] where the epoxy groups formed during oxidation were suggested to have a preference of aligning in a line. The aligned epoxy groups then induce a rupture of the underlying C-C bonds (Figure 1a).

However, it is still not clear how the graphene sheets can eventually break up, since even after the rupture of the C-C bonds the graphene sheet remains bridged by O atoms. Actually, a recent study shows that the mechanical strength of the graphene sheet is not strongly affected by the presence of epoxy chain and an epoxy line defect only weakens the fracture stress of the sheet by approximately 16%.[vii] This result indicates that, although the epoxy chain breaks the underlying C-C bonds, it does not really cause a breakup of the material by itself. The chemistry of the whole breakup process is still not clear.

Previous experimental study on GO has suggested the existence of carbonyl groups,[viii] and a very recent two dimensional NMR experiment shows that the carbonyl groups are spatially separated from the majority $sp^2$, C-OH, and epoxide carbons.[ix] This result indicates that carbonyl groups mainly distribute at the GO edge, and may thus be closely related to the oxidative break process. In this communication, based on density functional theory (DFT), we reveal how the oxygen attacks can break up atomic structure of graphene. Both the middle-site-initiated and the edge-site-initiated processes are studied, and the former based on intermediate epoxy pairs is found to be more favorable.

DFT calculations were performed with the Vienna Ab-initio Simulation Package (VASP)[x] using PBE exchange-correlation functional.[xi] The synchronous transit method with conjugated gradient refinements[xii] was used for transition state search.

Considering a graphene sheet with an epoxy chain already formed on it (Figure 1a), a natural question follows: are the O-connected carbon atoms more or less active? Since the epoxy chain induces a distortion from the ideal planar structure, we expect that the corresponding carbon atoms could become easier to be attacked during the further oxidation processes. Indeed, our calculations show that the energy of the epoxy-pair structure (Figure 1b) is 2.71 eV lower than that of the structure with the extra epoxy group located far from the epoxy chain. Therefore, once an epoxy chain appears, it is then energetically preferable to be further oxidized into epoxy pairs.

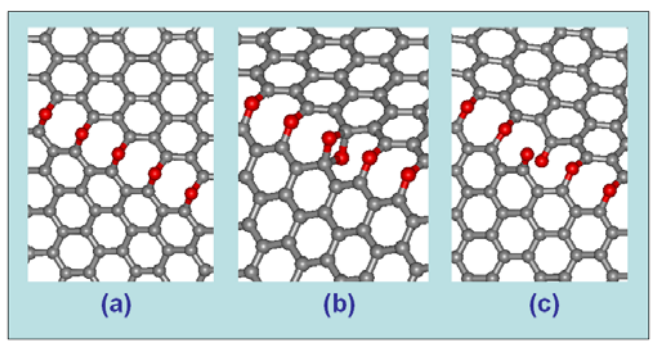

**Figure 1.** (a) A graphene sheet with an epoxy chain. (b) An epoxy pair or (c) a carbonyl pair is formed in the epoxy chain.

The additional energy gain by adding a new epoxy group to form an epoxy pair neighboring to an existing epoxy pair is 0.78 eV larger than the energy gain corresponding to forming an isolated epoxy group. Therefore, with the existing epoxy pairs, generating a new epoxy pair on the epoxy chain is still favorable in energy. We also note that, for a short epoxy chain, forming an epoxy pair or adding an epoxy group to extend the chain is comparable in energy (within 0.1 eV). This result suggests that epoxy pairs can form during the growth of the epoxy chain.

An epoxy pair is less stable than a carbonyl pair (Figure 1c), with a 0.48 eV difference in energy. It is difficult to form a carbonyl pair on a pristine graphene plane. A transition state search, however, gives an activation energy of 0.76 eV for the epoxy pair to carbonyl pair reaction (Figure 2a), and this value is lowered to 0.45 eV when there is a neighboring epoxy pair, which suggests a substantial reaction rate for the epoxy pair to carbonyl pair reaction at room temperature. Therefore, epoxy pair can be act as an intermediate species to form carbonyl pairs.

As expected, the energy barrier for concerted transition from two neighboring epoxy pairs to two carbonyl pairs is relatively high (1.07 eV). However, when the first epoxy pair to carbonyl pair reaction is completed, existing or newly formed neighboring epoxy pair can be quickly transferred to carbonyl pair with a small barrier of only 0.26 eV.

We have suggested here an atomic mechanism for oxidative graphene scission based on the new epoxy-pair species. However, in principle, under oxidation, carbonyl pairs can be directly formed by breaking the C-C bonds at a graphene edge. If such direct oxidative break of C-C bond can go inward, it leads to a new "tear-from-the-edge" mechanism.

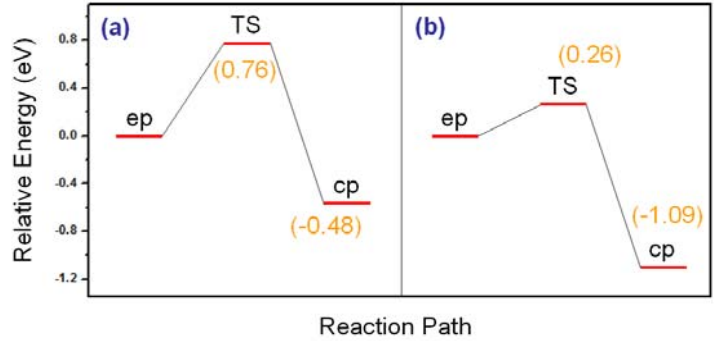

**Figure 2.** The relative energy of the reactant (epoxy pair, ep), transition state (TS), and product (carbonyl pair, cp) for (a) the first and (b) the next cp formation reactions. The numbers in parentheses are relative energies compared to reactants in eV.

To address this, we model a graphene edge using a super-cell ribbon structure (Figure 3a). As in slab model widely adopted for surface simulations, the coordinates of the carbon atoms on the left side are fixed to their two-dimensional-sheet values, and the next ten C columns are used to mimic a graphene edge. Dangling bonds on the left side are saturated by H atoms, while those at the right edge are terminated by hydroxy groups. A carbonyl pair bound to carbon atoms α and β is named cp(α, β), and we have studied the systems with a carbonyl pair cp(A6, A7) or cp(B7, B8). The former (Figure 3b) is more favorable in energy (1.08 eV). One possible reason is that the carbonyl pair cp(A6, A7) leads to two carboxyl groups and a hydrogen bond can be easily formed between them.

With the carbonyl pair cp(A6, A7) formed, what's the most favorable site for the next carbonyl pair? To quantitatively compare the preference between the go-inward and the attack-another-edge-bond processes, we calculate $\Delta E$ defined as

$$\Delta E = \frac{1}{2}(E_{cp2} + E_{pr}) - E_{cp} \qquad (1)$$

where $E_{cp2}$ is the energy of the unit cell with two aligned carbonyl pairs cp(A6, A7) and cp(C6, C7) (Figure 3c), $E_{cp}$ is the energy of the unit cell with cp(A6, A7), and $E_{pr}$ is the energy of the unit cell for the pristine OH-saturated graphene edge. Our calculations lead to a $\Delta E$ of 0.63 eV, indicating that new edge carbon bonds are easier to be attacked than those inside an existing carbonyl pair. Test calculations with a H-terminated edge also give a positive $\Delta E$ (0.45 eV). Since NMR experiment suggests that only part of the edge C atoms are oxidized to carboxyl groups,[xiii] it is highly unlikely the graphene sheet will be fully torn from the edge under oxidation.

Our cutting mechanism study thus suggests that epoxy pair is a critical intermediate species. We note that epoxy pairs may also relate to the "missing carbonyl" puzzle in GO structure study. In previous GO studies, a strong XPS peak around 289 eV was assigned to carbonyl groups.[viii,xiii] However, carbonyl exists mainly at edge sites,[ix,xiii] thus should not correspond to a strong peak. According to our model, epoxy pairs can widely exist on epoxy line defects. Since epoxy-pair bounded C and carbonyl C have similar chemical environment (both are bound with two valences to O), the former may be the dominant species contributing to the puzzling XPS peak. Actually, our atom in molecule (AIM) population analysis[xiv] assigns similar charge for carbonyl C (0.80) and those connecting to an epoxy pair (0.85), and both are much larger than the 0.43 positive charge assigned to C connecting to a single epoxy group.

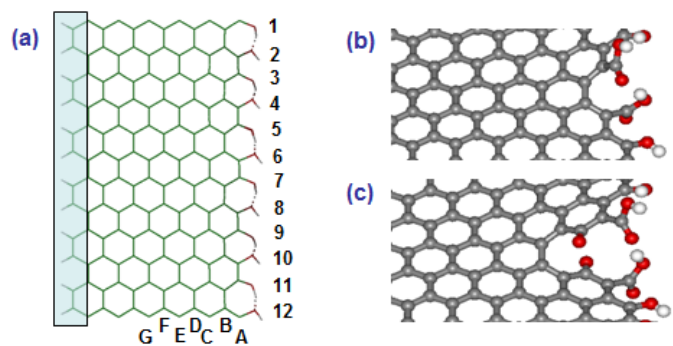

***Figure 3.*** (a) The geometric model to describe a graphene edge. Atoms in the shadow area were fixed during geometry optimization. (b) A carbonyl pair at the edge, and (c) another carbonyl pair formed one step inward.

In summary, insight into the graphene oxidative breakup process is obtained by first principles calculations, which is valuable for developing effective graphene manipulation means and for understanding the long puzzling GO structure. Our results suggest that a well controlled oxidation induced cut of graphene could lead to more smooth edges compared to heat or sonic treatment. In contrast to the previous unzipping mechanism, our new mechanism involves both sides of the graphene sheet. We expect a suppression of oxidative cut by, for example, transferring the oxidation reaction to a surface.

**Acknowledgment.** This work is partially supported by NSFC (20803071, 50721091, 20533030, 50731160010), by FANEDD (2007B23), by the National Key Basic Research Program (2006CB922004), by the USTC-HP HPC project, by the SCCAS and Shanghai Supercomputer Center.

**Supporting Information Available:** Computational details, optimized atomic coordinates, and absolute energies. This material is available free of charge via the Internet at http://pubs.acs.org.